\documentclass[12pt,preprint]{aastex}

\newcommand{\e}            {\mbox{$^{-1}$}}

\newcommand{\simgt}        {\gtrsim}
\newcommand{\simlt}        {\lesssim}
\newcommand{\pp}           {\noindent\hangindent 20pt\hangafter=1}
\def\n2hp{\mbox{N$_2$H$^+$}}
\def\c34s{\mbox{C$^{34}$S}}
\def\kms{\mbox{km s$^{-1}$}}

\def\plotfiddle#1#2#3#4#5#6#7{\centering \leavevmode
\vbox to#2{\rule{0pt}{#2}}
\includegraphics{#1}}

\received{August 2005}
\begin{document}

\title{A high resolution comparative study of the slowly contracting,
starless cores, L694-2 and L1544}
\author{Jonathan P. Williams}
\affil{Institute for Astronomy, 2680 Woodlawn Drive,
Honolulu, HI 96822; jpw@ifa.hawaii.edu}
\and
\author{Philip C. Myers}
\affil{Harvard--Smithsonian Center for Astrophysics,
60 Garden Street, Cambridge, MA 02138; pmyers@cfa.harvard.edu}

\shorttitle{Comparative study of L694-2 and L1544}
\shortauthors{Williams \& Myers}

\begin{abstract}
We present interferometric observations of \n2hp(1--0) in the starless,
dense core L694-2 and compare them to previously published maps of
L1544. Both cores are starless, centrally condensed, and show spectral
signatures of rotation and collapse.
We fit radially averaged spectra using a two-layer infall model
and measure the variation of opacity and infall speed in
each core. Both functions increase toward the center of each
core but the radial gradients are shallower, and the central
values lower, in L694-2.
This general behavior is predicted in models of gravitational
collapse with thermal plus magnetic support and the lower values
in L694-2 may be due to its lower mass or a slightly earlier
evolutionary state.
In either case, it appears that both cores will form stars
within a few $10^4$~yr.
\end{abstract}

\section{Introduction}
Low mass stars form via the gravitational collapse
of ``dense cores'' with densities $n({\rm H_2}) > 10^4$~cm$^{-3}$
which are built up either via the diffusion of neutrals past ions
tied to magnetic field lines or the dissipation of turbulent support
(Evans 1999). The Lynds catalog (Lynds 1962)
of optically selected dark cores provides an excellent starting
point for the observational study of the processes involved.
Lynds cores are generally compact, dense and emit strongly in a
wide range of molecular lines and continuum at millimeter wavelengths.
This allows for detailed investigations of their dynamics
and comparison with theoretical models (e.g. Gueth et al. 1997).

The first stage in star formation, the collapse of a starless
core, can be diagnosed through the analysis of self-absorbed
spectral features (Leung \& Brown 1977).
For sufficiently high optical depth and
a decreasing excitation temperature with radius, the outer
part of a core absorbs the emission from the center.
The resulting spectral profile will show a dip near the
velocity of the absorbing layer. If the core is collapsing,
this dip will appear at higher velocities than the bulk
of the core emission. Such red-shifted self-absorption is
the principal signature of inward motions.
Probably because self-absorption requires a fairly well-tuned
combination of optical depth and excitation conditions, however,
the window of observability of infall in starless cores is fairly short
(e.g. Gregersen \& Evans 2000).

L1544 is the prototypical example of a collapsing starless core.
It shows strong red-shifted self-absorption in a number of
molecular species (e.g., Tafalla et al. 1998; Caselli et al. 2002).
These have been interpreted to show both rotation and collapse
(Ohashi et al. 1999)
and the infall motions have been compared in detail
with a variety of collapse models,
ranging from gravitational free-fall (Whitworth \& Ward-Thompson 2001),
ambipolar diffusion (Ciolek \& Basu 2000),
external compression (Hennebelle et al. 2003),
and turbulent cooling (Myers \& Lazarian 1998).
L1544 has received so much attention as it shows the most clearly
defined infall signatures of any starless core.
It is important, nevertheless, to study other cores
to constrain the models further.

L694-2 was identified as a collapsing, starless core
by Lee and collaborators;
Lee \& Myers (1999) used the Digital Sky Survey to define
the optical properties of the most opaque Lynds cores
in detail and followed up with a molecular line survey
of those that were determined to be starless
(Lee, Myers, \& Tafalla 1999).
L694-2 came to prominence in this work as a
``strong infall candidate'' on the basis of pronounced
red-shifted self-absorption in the CS(2--1) line.
Subsequently Lee, Myers, \& Plume (2004) showed similar infall
profiiles in CS(3--2) and DCO$^+$(2--1). Their spectral fits
indicated that the infall speed increased with optical
depth and, presumably therefore, toward the center of the core.

The starless nature of L694-2 was deduced from the lack of an IRAS
point source and confirmed by a VLA non-detection at 3.6~cm 
(Harvey et al. 2002). Together these limit the luminosity
of an embedded source to $L\simlt 0.3~L_\odot$.
Subsequent millimeter continuum observations by Harvey et al. (2003a)
showed that there was no compact, disk-like source to a sensitivity
$M_{disk}\simlt 5\times 10^{-4}~M_\odot$.
The core is included in the ``c2d'' Spitzer Legacy program
which will show whether a very low luminosity source exists
(Evans et al. 2003)

L694-2 also has similar structural and chemical properties as L1544
(Crapsi et al. 2005).
The low temperatures and high densities of starless cores result in
substantial molecular depletion onto grain surfaces (Tafalla et al. 2002).
These chemical properties imply that to follow the dynamics toward
the core centers requires a relatively inert molecular probe that remains
in the gas phase. To measure relative (infall) motions, in particular,
requires high optical depth and therefore high abundance.
Only a handful of molecules meet these conditions.

\n2hp\ is well suited as a relatively abundant species
in starless cores that suffers little depletion except at
the very highest column densities (Bergin et al. 2002).
Tafalla et al. (1998) found that the $J=1-0$ line was double-peaked
due to self-absorption at the center of L1544.
Williams et al. (1999; hereafter Paper I) subsequently mapped
this line at $10''$ resolution to examine the velocity field
at scales approaching that of a protostellar disk (1000~AU).
In that work, we found a centrally condensed, slightly elongated
core, and an increasing infall velocity toward the center.

Lee, Myers \& Tafalla (2001) mapped \n2hp(1--0)
emission in L694-2 at $50''$ resolution and found
a compact, centrally condensed core.
Here, we present combined interferometric plus singledish
observations of \n2hp\ in L694-2 at $10''$ resolution.
Although the \n2hp\ spectra do not split into two
distinct peaks as in L1544, they do have identifiable
``red shoulders'' due to self-absorption that are equally indicative
of collapse. We compare the two datasets and model the spectra in
both cores using new analytic tools made available by
De Vries \& Myers (2005).

The observations are presented in \S2, and the results and modeling
in \S3. We show that L1544 is more centrally condensed and has a higher
infall speed and gradient than L694-2. This general trend is
discussed in the context of gravitational collapse models in \S4.
Throughout the paper, we adopt a distance to L694-2 of 230~pc
based on optical star counts by Kawamura et al. (2001)
and 140~pc for L1544 based on its association with the Taurus
cloud (Elias 1978).

\section{Observations}
L694-2 was observed with the 10 element Berkeley-Illinois-Maryland
array\footnotemark\footnotetext{The BIMA array is operated with support
from the National Science Foundation under grants AST-9981308 to
UC Berkeley, AST-9981363 to U. Illinois, and AST-9981289 to U. Maryland.}
in its compact C configuration on 1999, June 7, 8 and 11.
The phase center of the observations was
$\alpha(2000)=19^{\rm h} 41^{\rm m} 04^{\rm s}.5$, 
$\delta(2000)=10^\circ 57' 02''$
and the total (on-source) integration time was 14 hours.
The amplitude and phase calibration was performed from 3 minute
integrations of the quasar 1925+211 interleaved with each 22
minute observation on source. The passband and flux calibration
was carried out from a 30 minute observation of Uranus toward
the end of each track. The derived flux for 1925+211 was 1.7~Jy.

The digital correlator was configured with 512 channel windows
at a bandwidth of 6.25~MHz ($0.04$~\kms\ velocity resolution per channel)
centered on the 7 hyperfine components of \n2hp(1--0)
in the lower sideband and \c34s(2--1) in the upper sideband.
Eight 32 channel windows at a bandwidth of 100~MHz.
were used to measure continuum radiation.
Data were reduced with the MIRIAD package using standard procedures.
The data sets from each day were calibrated separately and then
transformed together with natural weighting.
The resulting beamsize was $9\farcs 5 \times 7\farcs 1$.

The \n2hp\ lines were detected with a high signal-to-noise ratio
in the lower sideband, but no emission was detected in the upper
sideband, implying a $3\sigma$ upper limit of 1.5~K~\kms\ for
the \c34s\ emission integrated over velocities $9.3-9.9$~\kms.
No continuum emission was detected at a $3\sigma$ level of
2.1~mJy~beam\e.

The projected baselines of the interferometer data ranged
from 1.9 to 27~k$\lambda$. To provide information at lower
spatial frequencies, we observed L694-2 using the 
Five College Radio Astronomy Observatory 14~m telescope
(FCRAO\footnotemark\footnotetext{FCRAO is supported in part by the
National Science Foundation under grant AST-0100793 and is operated
with permission of the Metropolitan District Commission, Commonwealth
of Massachusetts}) on 1999 December 20.
We used the 16 element, focal plane array SEQUOIA and a system of
autocorrelation spectrometers at a spectral resolution of 20~kHz.
Standard position-switching procedures were used and resulted in a
Nyquist sampled map with an rms noise of 0.05~K per 0.06~\kms\ channel.
Antenna temperatures were multiplied by a gain of 43.7~Jy~K\e\
to convert to flux units and the singledish data were then added
to the interferometer data using the combination method described in
Stanimirovic et al. (1999).
The analysis in this paper was performed on the combined
BIMA+FCRAO dataset convolved to a spatial resolution of $10''$
and interpolated to velocity channels of width 0.05~\kms.

L1544 had been previously observed in a similar way with the BIMA
array and IRAM 30~m antenna (Paper I).
To make an unbiased comparison between the two cores,
we re-reduced the data and combined the interferometer and
singledish datasets using the same linear combination technique
described above.

\section{Results}
\subsection{Spectral comparison}

The $J=1-0$ transition of \n2hp\ has 7 hyperfine components, two groups
of three and one isolated (Caselli, Myers \& Thaddeus 1995).
The linewidths of L694-2 and L1544 are similar and sufficiently
small that all 7 components can be distinguished.
Figure~1 shows three different hyperfine components,
ordered from high to low optical depth, toward the two cores.
The L1544 data have been smoothed to $16''$ to achieve the same
linear resolution, 2300~AU, as the L694 data.
In each case, the spectra are most asymmetric in the
top panel and become more symmetric as the optical depth of the
hyperfine component decreases in the middle and bottom panels.
The asymmetries must be due, therefore, to radiative transfer
effects rather than kinematics. We conclude that both cores
exhibit red-shifted self-absorption. This is strong evidence
for core collapse.

The self-absorption features are less pronounced in the L694-2
spectra, which show shoulders, than in the L1544 spectra,
which show dips. This is indicative of a lower opacity
and is in agreement with the multi-transition analysis by
Crapsi et al. (2005) who found that the \n2hp\ column density
is $\sim 25-30\%$ lower in L694-2 than in L1544.

The isolated component, $F_1 F=10-11$, has the lowest optical
depth and therefore is the weakest and most symmetric.
Although we were able to analyze the dynamics using this line
in L1544 in Paper I, the lower optical depth and
signal-to-noise ratio precludes a similar study
of the L694-2 dataset. We therefore focus the remainder of
the analysis and comparison of the two cores on the central
$F_1 F=23-12$ component at 93.1737767~GHz, as it is the
brightest and has the highest optical depth, making it the
most sensitive for tracing relative motions in the core.

\subsection{Moment maps}

Maps of the first two moments of the
$F_1 F=23-12$ hyperfine component are plotted
in Figures~2 (L694-2) and 3 (L1544).
The contours of zeroth moment, or integrated intensity,
are similar to dust continuum images
(Visser et al. 2002; Ward-Thompson, Motte, \& Andr\'e 1999)
suggesting that there is negligible \n2hp\ depletion in each core and
that this transition is a reliable tracer of the dynamics toward
the core centers at $\sim 2000$~AU resolution scales.

The first moment, or mean velocity, of the line is
overlaid in color scale over the integrated intensity contours.
In L694-2, there is a central low velocity (blue)
hole surrounded, almost in a complete ring, by higher velocities.
In L1544, there is a clear trend of low to high (red) velocities
across the core, but the lines of constant velocity are curved
in a `C'-shape. In each case, the velocity pattern cannot be
explained solely by rotation or shear across the core.
Rather, following the earlier work by Tafalla et al. (1998)
in L1544 and Lee et al. (1999) in L692-2,
we suggest that the \n2hp\ velocity pattern
can be explained by a combination of rotation and core contraction.

The spectroscopic signature of core collapse is red-shifted
self-absorption. The mean velocity of such spectra is lower than in
the absence of self-absorption or infall and therefore appears
in a first moment map as a dip toward low, or blue, velocities.
Core rotation manifests itself as an one-dimensional gradient in
the first moment map, however, and the two effects can therefore
be distinguished (Adelson \& Leung 1988).
Walker, Narayanan, \& Boss (1994) modeled the radiative transfer
of collapsing, rotating cores and coined the phrase ``blue-bulge''
to describe the feature in the mean velocity field
where infall motions dominate over rotation. For a centrally
condensed core with uniform rotation and increasing infall
speed toward the center, the contours of constant velocity form
a 'C'-shape near the peak of emission, dependent on the
inclination of the rotation axis to the line of sight.
They point out that this feature is a more robust signature of
infall than the classic line asymmetry and subsequently
demonstrate its applicability in two star-forming systems
(Narayan, Walker, \& Buckley 1998; Narayan et al. 2002).

The mean velocity maps of both L694-2 and L1544 also show
the ``blue-bulge'' signature of collapse superimposed on
rotation. A central blue dip can be clearly seen toward
the center of L694-2. The velocity gradient is higher
in L1544 so the lines of constant velocity do not wrap
onto themselves but nevertheless are clearly curved.
Using these maps, we are able to follow the core dyanamics beyond
the center where the spectral asymmetries in Figure~1 are most clear.

To isolate and analyze the infall motions, we removed the rotation
in each core by subtracting a linear velocity gradient from the maps.
This is an approximation as the \n2hp\ line is optically thick
and self-absorbed. We define the best fit, however, by minimizing the
dispersion in mean velocity within the central $40''$,
and it is weighted (simply by the number of pixels)
to the outer parts of the core where the optical depth is low.
Therefore, the error in measuring the rotation is likely to be small.
The subtracted mean velocity image are shown for L694-2 and
L1544 in the right panels of Figures 2 and 3 respectively.
The magnitude and direction of the rotation vector is
shown in the bottom right corner. The velocity gradient in
L1544, 4.1~\kms, is similar to the value determined in Paper I
from a different modeling technique.
The gradient in L694-2 is much smaller, 0.5~\kms.
Without knowledge of the core geometries and inclination of
the rotation axes, however, it is not possible to compare
their angular momenta directly.

In both cores, the residual (mean minus rotation) velocity field
shows a marked bipolar deviation from circular symmetry.
The signal-to-noise ratios of the data are
unfortunately too low to study this quantitatively but it is
perhaps noteworthy that the bipolarity is not aligned with
the direction of core elongation.
This suggests that the infall speed is not spherically symmetric,
as would be expected for a purely gravitational flow resisted by
isotropic pressure.

The high velocity gradient in L1544 stretches the color scale
and masks the central blue-bulge feature in the residual map.
The increase in the residual velocity field away from the
core center is shown for both cores in Figure~4 and demonstrates
that the bluebulge dip is steeper in L1544 than L694-2.
This could be due to either or both a steeper radial variation
of opacity and infall speed. To examine this further we return to
an examination of the spectral profiles.

\subsection{Infall modeling}

By removing the velocity field due to rotation, we can
isolate the infall dynamics in each core.
Away from the center, however, the signal-to-noise ratios
are too low to allow modeling of individual spectra and we
binned the data into concentric radial annuli with $10''$ width.
This allows us to measure the radial variations of optical depth
and infall speed. The radial averaging is justified in L694-2 by the
approximate circular symmetry of the core and the analysis by
Harvey et al. (2003a, 2003b) who showed that the core density profile
can be fit either by a spherically symmetric model or a cylinder that
is nearly pole on. Harvey et al. (2003b) also noted that a cylindrical
configuration would be prone to collapse. L1544 is more elongated
but we found an approximately circular distribution of infall
speed in Paper~I.

The average spectra for the two cores are shown as
histograms in Figure~5. The systemic motion of each core
(defined from the fits described below) is shown as a dotted
vertical line. Moving out from the core centers (top panels)
to the outer parts (lower panels), the spectra in both cores become
weaker, more symmetric, and their mean velocity shifts to higher values.
This behavior can be explained by an absorbing layer of gas falling
onto the core. At small radii, the optical depths are high and the
profiles show relatively strong self-absorption.
Because the absorbing layer is falling onto the core, the
absorption is redshifted with respect to our line of sight and
the profiles therefore appear asymmetric.
As the optical depth decreases at larger radii, there is
less self-absorption and the spectra become more symmetric.

The values of the optical depth and infall speed can be
determined via detailed modeling.
We were able to produce very good fits to the radially
averaged spectra using the two layer infall model described
in Myers et al. (1995) and the parameter search technique
of De Vries \& Myers (2005). Following the recommendation in
that paper, we used the {\tt hill5} De Vries \& Myers model.
The {\tt hill5} analytic models are characterized by 5 parameters:
the systemic velocity, velocity dispersion, total optical depth,
peak temperature, and the infall speed; $v_{LSR}, \sigma, \tau,
T_{\rm peak}$, and $v_{in}$ respectively.
The infall speed is defined such that the front layer moves at
velocity $v_{LSR}+v_{in}$ with respect to the observer and the rear
layer at velocity $v_{LSR}-v_{in}$.
Measurement of the infall speed requires a sufficiently high
opacity to create significant self-absorption in the spectra,
and was not possible in the outer parts of each core.
The fits are only shown, therefore, for the spectra
where the modeled optical depth is greater than unity;
$\theta < 30''$ for L694-2 and $\theta < 40''$ for L1544.
The model spectra are overlaid as solid lines over the
histograms in Figure~5 and the parameters that define them
are listed in Table~1.

The parameter uncertainties were estimated by taking the
best fit models in Table~1, adding random noise with the
same rms as the observed noise, and re-fitting using the same
procedure. The error in each parameter was set equal to the
dispersion in the fits over 50 repeats of this process.
We found that the errors in $v_{LSR}$, $\sigma$, and $T_{\rm peak}$
were negligible, but the errors in $\tau$ and $v_{in}$,
whose value depends on relatively subtle features in the
spectral profiles, were larger and ranged from $\sim 5-15\%$.

The radial variations of modeled optical depth and infall speed
are shown in Figure~6. Within the $\sim 0.01$~\kms\ uncertainties,
the analysis here indicates similar, but slightly higher,
infall velocities in L1544 than those measured in Paper I
using a different fitting method to the isolated hyperfine component.
We can be more confident of the relative differences between
L1544 and L694-2, however, because we have applied an identical
analysis to observations of the same hyperfine component.
We find that both the optical depth and infall speed
are higher in the central 3500~AU in L1544 than in L694-2 and
both decline more steeply with radius.
The higher opacity was immediately apparent from the
stronger self-absorption features in the spectra in Figure~1.
The rapid decline toward larger radii reflects the steeper
blue-bulge signature seen in Figure~4.

\section{Discussion and Conclusions}

We have compared similar maps of the $J=1-0$ transition of \n2hp\
in two collapsing, starless cores, L694-2 and L1544.
Toward the core centers, the hyperfine components show an increasing
asymmetry with optical depth. This is strong evidence for inward
motions and complements earlier studies of collapse at larger
scales by Lee et al. (2001) in L694-2 and Tafalla et al. (1998)
in L1544, using lower resolution observations of CS.

We then analyzed the dynamical structure of the cores in
more detail by examining the moment maps of the most
optically thick, $F_1F=23-12$ hyperfine component.
The mean velocity pattern showed a systematic variation
across the cores from rotation and an additional
``bluebulge'' toward the center from infall.
This follows earlier modeling work and observations of other
star forming regions by Walker et al. (1994) and Narayanan (1997).
By removing the velocity gradient from the data, we were able
to isolate the blueward shift due to infall asymmetry.
The depth of the bluebulge is similar, $\sim 0.05$~\kms,
in both cores but the size of the dip is larger
in L694-2 reflecting a shallower opacity gradient.
These velocity-shifted spectra were then radially averaged and
modeled using an automated parameter search technique by
De Vries \& Myers (2005) applied to the two-layer infall
model of Myers et al. (1995).

The spectral fits show that both cores have low velocity dispersions,
$\sigma\sim 0.08~\kms\ = 1.5\sigma_T(\n2hp)$,
for a kinetic temperature $T_K=10$~K.
The non-thermal motions are therefore very small,
$\sigma_{NT}/\sigma_T({\rm H_2})\simeq 1/3$.
In fact, much of the non-thermal motion may be due to infall itself.
Whether this is the case or not, the pressure support in the cores,
which is proportional to the square of the velocity dispersion,
is predominantly thermal. The turbulence that is seen at larger
scales has almost completely dissipated and the infall motions that
we have measured here can constrain theoretical models of the
gravitational flows in a core supported by a thermal, or thermal
plus magnetic, pressure (e.g. Shu 1977; Mouschovias 1976).

Indeed the variation of infall speed with radius in L1544 has been
successfully reproduced using an ambipolar diffusion model
by Ciolek \& Basu (2000)
and the free-fall collapse of a sphere with Plummer-like
density profile by Whitworth \& Ward-Thompson (2001).
These two models differ in that the total star formation timescale
is very different, $t_{\rm sf}=2.7$~Myr for ambipolar diffusion,
compared to $t_{\rm sf}=0.04$~Myr for the free-fall collapse
timescale of the central flat region of the Plummer core.
They agree, however, in how far L1544 is from the end of this process.
Ambipolar diffusion lengthens the build-up of a core to a
critical state where magnetically diluted collapse takes place
and observable infall motions only occur in the last few
percent of $t_{\rm sf}$. The best fit Ciolek \& Basu model
places the age of L1544 at $t\simeq 0.99t_{\rm sf}$, or only
$3\times 10^4$~yr from forming a star.
Whitworth \& Ward-Thompson (2001) place the age of L1544 
of about 50\% of the central free-fall tiemscale so
only $2\times 10^4$~yr from star formation.
Myers (2005) also predicts a similar late stage of collapse
for L1544 in a purely gravitational collapse model
applied to different geometries.

A general feature of both models (and indeed any model of gravitational
collapse) is that the infall speed increases with time {\it at all radii}.
The central density also increases, although the density profile in
the outer parts, radii $\simgt 2000$~AU, is independent of time.
At the resolution of these observations, the density profiles of
the two cores should therefore appear very similar but the central
column density should be higher in the more evolved core
and the infall speeds should be higher throughout.

The main results of the spectral fitting are the radial variations of
infall speed and opacity in the two cores (Table~1 and Figure~6).
In the inner regions, $r < 3500$~AU, L1544 is $\sim 25\%$ more opaque
and collapsing about $\sim 15\%$ faster.
This may indicate that L694-2 is somewhat less evolved toward star
formation than L1544, but it may also be due to its lower mass.
The precise masses of the cores depend on how their outer boundaries
are defined, but within the $\sim 10^4$~AU size of our \n2hp\ maps,
we estimate a mass $\sim 1~M_\odot$ for L694-2 from Visser et al. (2002)
and $\sim 1.5~M_\odot$ for L1544 from Ward-Thompson et al. (1999).
Whether evolution or mass explains the differences in the model fits,
the centrally concentrated blue-bulge signatures of collapse are
remarkably similar and indicate that point mass formation at the
centers of {\it both} cores will occur in a few $10^4$~yr.

We caution that the observed values of the radii in the plot
of infall speed (Figure~6) do not directly translate to physical
core radii because the spectral features essentially convolve the
emission from several distinct regions along the line of sight.
A quantitative comparison with theory requires calculating the
radiative transfer of dynamical models (e.g. Choi et al. 1995).
This is also necessary to tie the \n2hp\ observations presented
here to measurements of infall at large scales from CS observations
(Lee et al. 2001; Tafalla et al. 1998).
This adds the significant additional complexity of core chemistry.
The relevant chemical timescales for molecule formation and
depletion are similar to the dynamical timescales of collapse.
Consequently, the comparison of infall motions measured in two
different molecular species requires combined modeling of
structure, dynamics, chemistry, and radiative transfer
(e.g. Doty, Sch\"oier, \& van Dishoeck 2004).

Due to the relatively low signal-to-noise ratio of these
data and the subtle self-absorption features in the spectra,
we have only been able to examine the radial variation
of core properties. The residual moment maps, after subtraction
of a a velocity gradient, show additional structure.
These approximately bipolar velocity features are not
aligned with the elongation of the cores, measured from
the central intensity contours of L694-2 and L1544.
Optical depth effects alone, therefore, are not the sole
explanation and it appears that the infall speed is not
spherically symmetric.
Both core elongation and a preferred direction for collapse
are expected from a static magnetic field geometry.
A quantitative investigation of the two-dimensional
density and velocity structure requires more sophisticated
radiative transfer models (e.g. Hogerheijde, \& van der Tak 2000)
and higher quality data, such as will soon be available with
the CARMA array (Woody et al. 2004).

Several other starless collapsing core candidates are known.
It will be very worthwhile to study them at high resolution in
molecular species that probe the dynamics at their centers.
The comparison of such data will give further insight into the
processes that collapse the dense ISM down to planetary disk scales.

\acknowledgments
We thank Chris De Vries for advice concerning his code
to fit the spectra.
We acknowledge support from NSF grant AST-0324328 (JPW)
and NASA Origins of Solar Systems grant NAG5-13050 (PCM).

\section{References}
\parskip=0pt
\bigskip

\pp Adelson, L.~M., \& Leung, C.~M.\ 1988, \mnras, 235, 349 

\pp Bergin, E. A., Alves, J., Huard, T., \& Lada, C. J. 2002, ApJ, 570, L101


\pp Caselli, P., Walmsley, C. M., Zucconi, A., Tafalla, M., Dore, L.,
    \& Myers, P. C. 2002, ApJ, 565, 331

\pp Caselli, P., Myers P.C., \& Thaddeus, P. 1995, ApJ, 455, L77

\pp Choi, M., Evans, N.~J., Gregersen, E.~M., \& Wang, Y.\
    1995, \apj, 448, 742

\pp Ciolek, G. E., \& Basu, S. 2000, ApJ, 529, 925

\pp Crapsi, A., Caselli, P., Walmsley, C. M., Myers, P. C., Tafalla, M.,
    Lee, C. W., \& Bourke, T. L. 2005, ApJ, 619, 379


\pp De Vries, C.~H., \& Myers, P.~C.\ 2005, \apj, 620, 800 

\pp Doty, S.~D., Sch{\" o}ier, F.~L., \& van Dishoeck, E.~F.\
    2004, \aap, 418, 1021 

\pp Elias, J.~H.\ 1978, \apj, 224, 857 

\pp Evans, N. J. et al. 2003, PASP, 115, 965

\pp Evans, N. J. 1999, ARAA, 37, 311

\pp Gregersen, E.~M., \& Evans, N.~J.\ 2000, \apj, 538, 260 

\pp Gueth, F., Guilloteau, S., Dutrey, A., \& Bachiller, R.\
    1997, \aap, 323, 943 

\pp Harvey, D. W. A., Wilner, D. J., Di Francesco, J., Lee, C. W.,
    Myers, P. C. \& Williams, J. P. 2002, AJ, 123, 3325

\pp Harvey, D. W. A., Wilner, D. J., Myers, P. C.,
    \& Tafalla, M. 2003a, ApJ, 597, 424

\pp Harvey, D. W. A., Wilner, D. J., Lada, C. J., Myers, P. C.,
    \& Alves, J. F. 2003b, ApJ, 598, 1112

\pp Hennebelle, P., Whitworth, A.~P., Gladwin, P.~P., \& Andr{\' e}, P.\
    2003, \mnras, 340, 870 

\pp Hogerheijde, M.~R., \& van der Tak, F.~F.~S.\ 2000, \aap, 362, 697 

\pp Leung, C.~M., \& Brown, R.~L.\ 1977, \apjl, 214, L73 

\pp Kawamura, A., Kun, M., Onishi, T., Vavrek, R., Domsa, I., Mizuno, A.,
    \& Fukui, Y. 2001, PASJ, 53, 1097

\pp Lee, C. W., \& Myers, P. C. 1999, ApJS, 123, 233

\pp Lee, C. W., Myers, P. C., \& Tafalla, M. 1999, ApJ, 526, 788

\pp Lee, C. W., Myers, P. C., \& Tafalla, M. 2001, ApJS, 136, 703

\pp Lee, C. W., Myers, P. C., \& Plume, R., 2004, ApJS, 153, 523

\pp Lynds, B. T. 1962, ApJS, 7, 1

\pp Mouschovias, T.~C.\ 1976, \apj, 207, 141 

\pp Myers, P.~C.\ 2005, \apj, 623, 280 

\pp Myers, P. C., \& Lazarian, A. 1998, ApJ, 507, L57

\pp Narayanan, G. 1997, PhD thesis, University of Arizona

\pp Narayanan, G., Moriarty-Schieven, G., Walker, C.~K., \& Butner, H.~M.\
    2002, \apj, 565, 319 

\pp Narayanan, G., Walker, C.~K., \& Buckley, H.~D.\ 1998, \apj, 496, 292 

\pp Ohashi, N., Lee, S.~W., Wilner, D.~J., \& Hayashi, M.\ 1999,
    \apjl, 518, L41 


\pp Shu, F.~H.\ 1977, \apj, 214, 488

\pp Stanimirovic, S., Staveley-Smith, L., Dickey, J. M., Sault, R. J.,
    Snowden, S. L. 1999, MNRAS, 302, 417


\pp Tafalla, M., Myers, P.~C., Caselli, P., Walmsley, C.~M.,
    \& Comito, C.\ 2002, \apj, 569, 815

\pp Tafalla, M., Mardones, D., Myers, P.C., Caselli, P., Bachiller, R.,
    \& Benson, P.J. 1998, ApJ, 504, 900

\pp Visser, A.~E., Richer, J.~S., \& Chandler, C.~J.\ 2002, \aj, 124, 2756 

\pp Ward-Thompson, D., Motte, F., \& Andre, P.\ 1999, \mnras, 305, 143 

\pp Walker, C. K., Narayanan, G., \& Boss, A. P. 1994, ApJ, 431, 767

\pp Whitworth, A.~P., \& Ward-Thompson, D.\ 2001, \apj, 547, 317 

\pp Williams, J.P., Myers, P.C., Wilner, D.J., \& Di Francesco, J.
    1999, ApJ, 513, L61 (Paper I)

\pp Woody, D.~P., et al.\ 2004, \procspie, 5498, 30

\clearpage
\begin{figure}[ht]
\plotfiddle{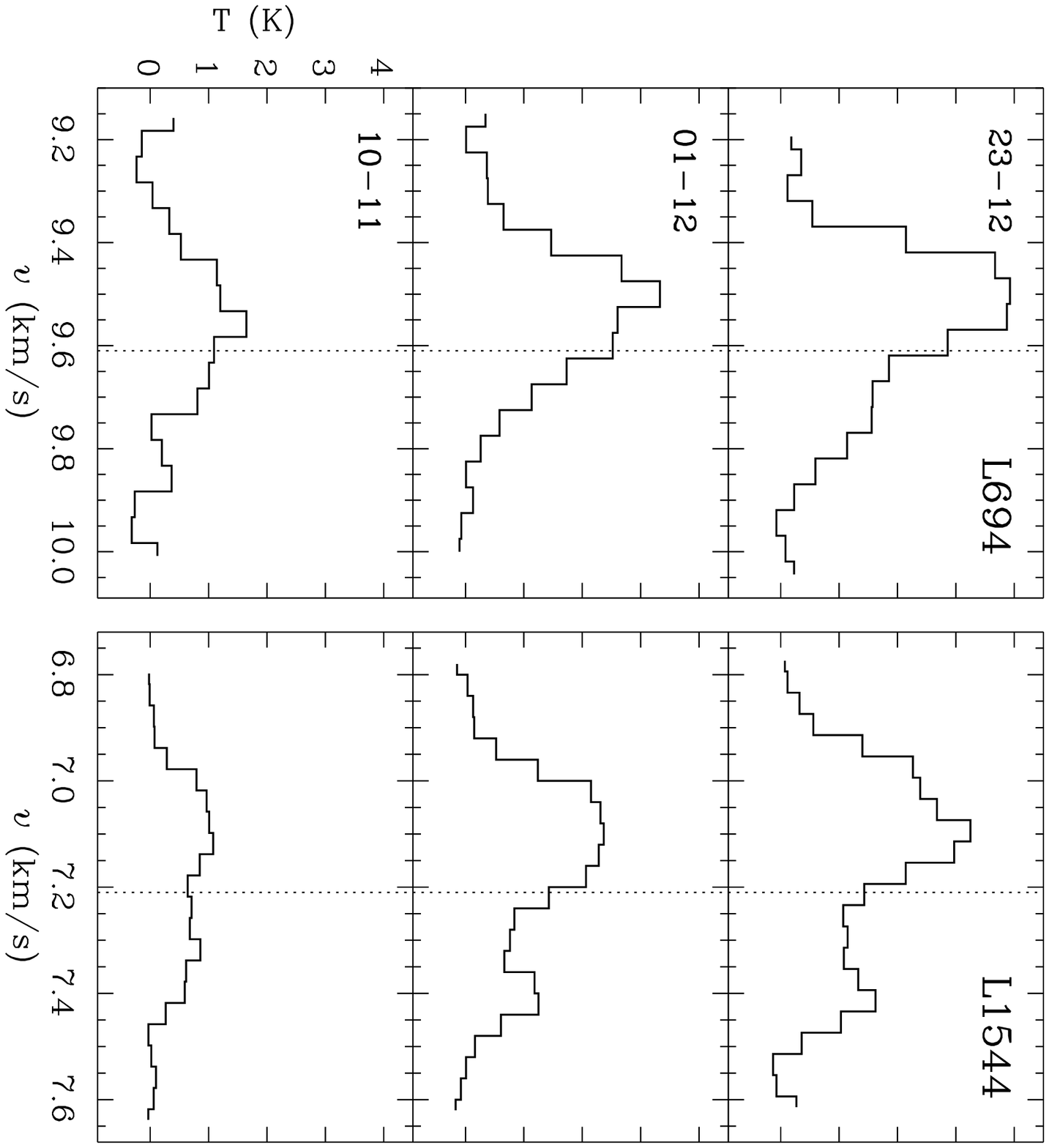}{50pt}{90}{70}{70}{280}{-360}
\end{figure}
\vskip 4.6in
\noindent{\bf Figure 1:}
Comparison of \n2hp(1--0) spectra toward the centers of L694-2 and L1544
at the same linear resolution, 2300~AU. Three hyperfine components are
shown, ordered from high to low optical depth, and the dotted line shows
the systemic velocity for each core. The increasing asymmetry with
opacity in each core is strong evidence for collapse.

\clearpage
\begin{figure}[ht]
\plotfiddle{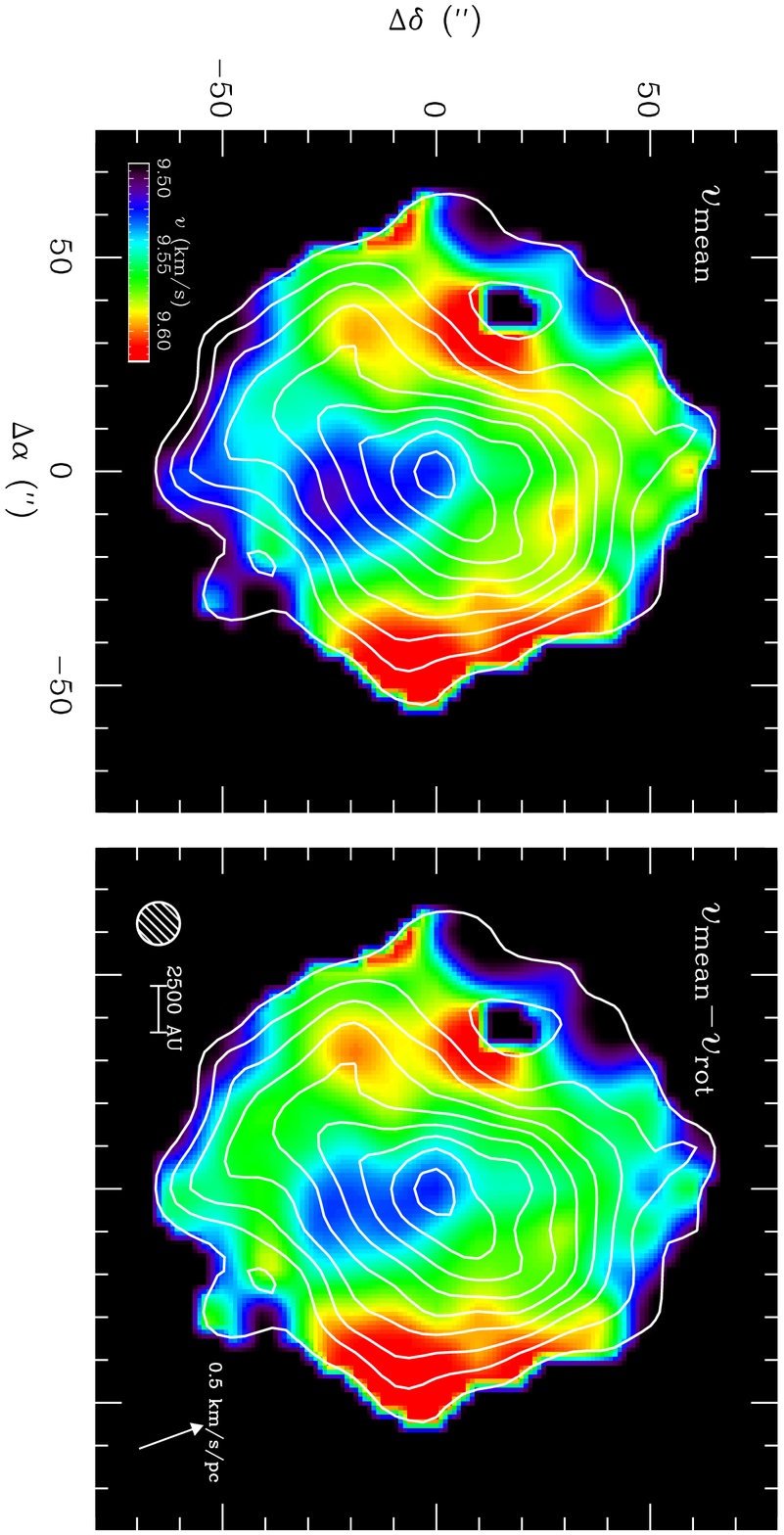}{50pt}{90}{60}{60}{240}{-250}
\end{figure}
\vskip 2.9in
\noindent{\bf Figure 2:}
\n2hp\ $1_{23}-0_{12}$ intensity and velocity structure in L694-2.
The left panel shows the mean velocity in colors overlaid on contours
of integrated emission.
The right panel shows the residual velocity field after subtracting
a linear gradient due to core rotation. The direction and magnitude of
the velocity gradient are shown in the lower right corner.
The $10''$ resolution of the data and scale bar are shown in the
lower left corner. The color scale ranges from 9.49 to 9.62~\kms\
and contours have starting point and increments of 0.12~K~\kms\
in both panels.

\clearpage
\begin{figure}[ht]
\plotfiddle{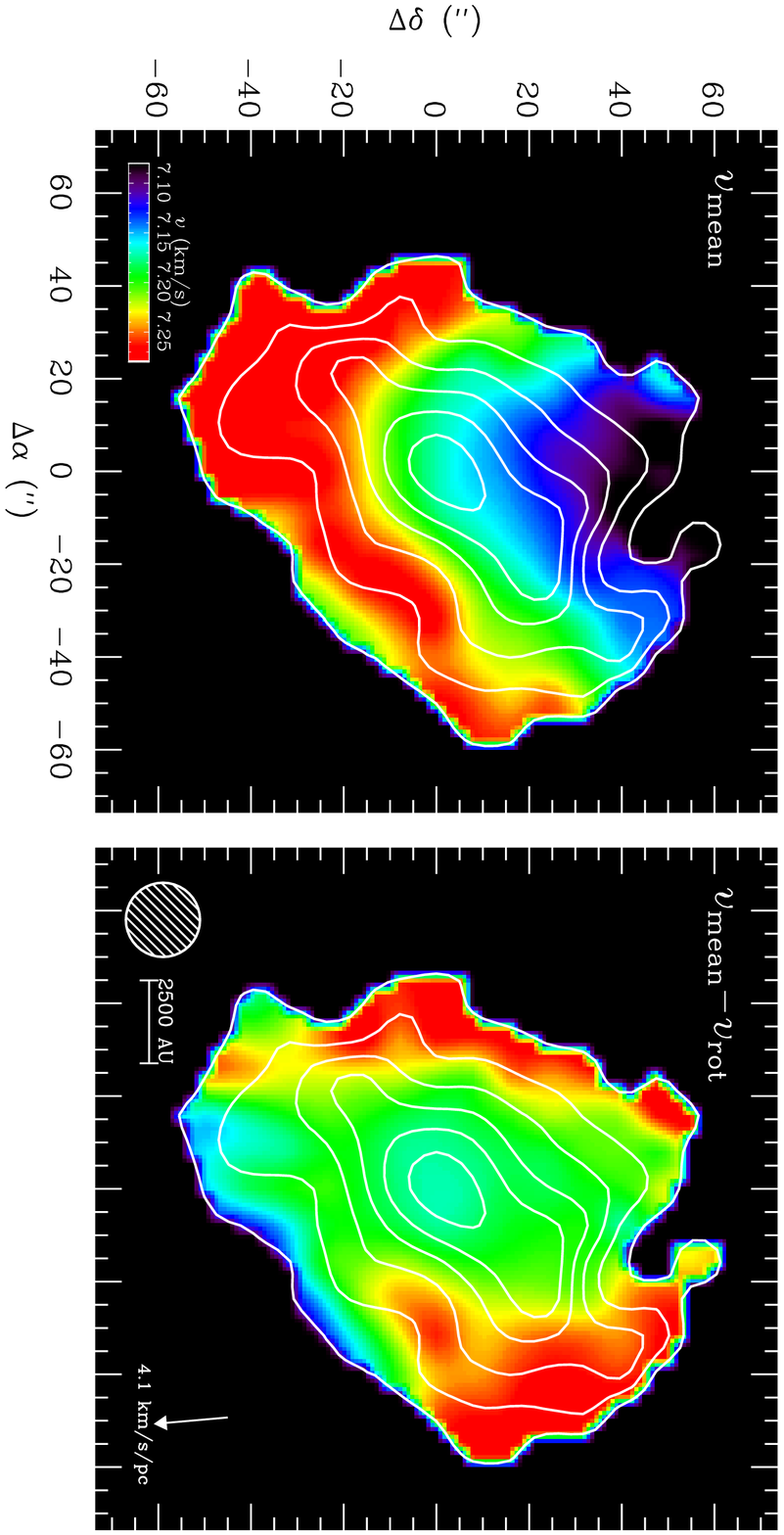}{50pt}{90}{60}{60}{240}{-250}
\end{figure}
\vskip 2.9in
\noindent{\bf Figure 3:}
Similar plot as Figure~1 for L1544.
The color scale ranges from 7.08 to 7.28~\kms\
and contours have starting point and increments of 0.16~K~\kms\
in both panels. These plots were made with the data smoothed to
the same linear resolution, 2300~AU, as the L694-2 data.
Note that the velocity gradient is much higher in L1544
and the color scale is therefore stretched over a wider
range. This has the effect of reducing the contrast
of the central blue-bulge feature.

\clearpage
\begin{figure}[ht]
\plotfiddle{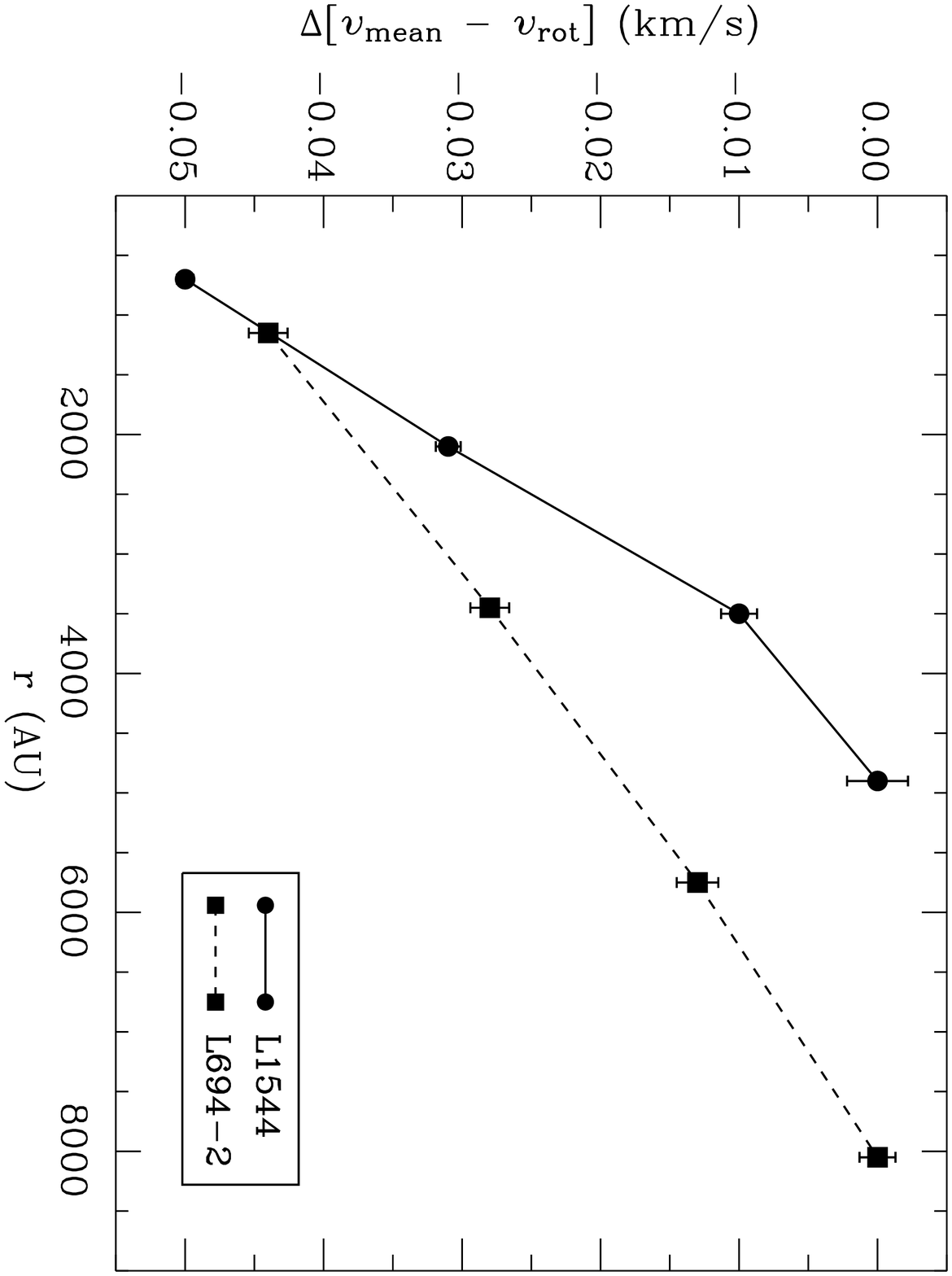}{50pt}{90}{60}{60}{235}{-280}
\end{figure}
\vskip 3.5in
\noindent{\bf Figure 4:}
Comparison of the blue-bulge signature in L694-2 and L1544.
Each point is a radial average in $10''$ annuli of the residual velocity
field in the right panels of Figures~2 and 3. The velocity scale
is set to zero at the outermost radii to show the blueward shift
of mean velocity toward the center of each core.
The errors are the standard deviation of the mean.
The bluebulge is larger in extent but shallower in L694-2
(squares connected by a dashed line) compared to L1544
(circles connected by a solid line).

\clearpage
\begin{figure}[ht]
\plotfiddle{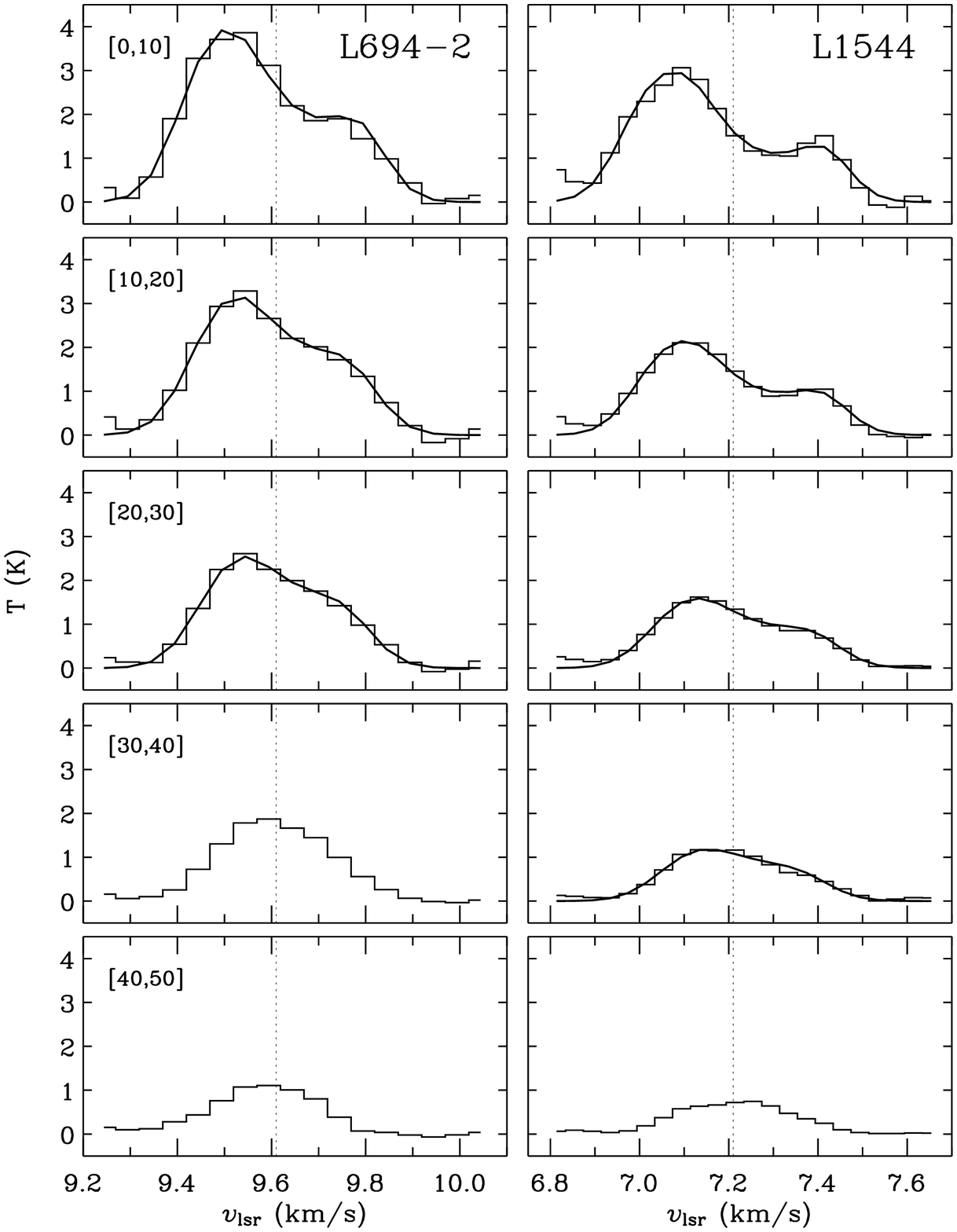}{200pt}{0}{60}{60}{-190}{-250}
\end{figure}
\vskip 3.3in
\noindent{\bf Figure 5:}
Radially averaged \n2hp\ $1_{23}-0_{12}$ spectra (histogram)
and two-layer model fits (line) for the two cores.
L694-2 is on the left hand side, L1544 on the right.
The averaging interval, in arcseconds, is shown in the top left
hand corner of the left panel and is the same in the right panel.
The vertical dotted line is the systemic velocity of each core.
The shoulder like profile toward the center of L694-2
contrasts with the two distinct peaks toward the center
of the more opaque L1544.

\clearpage
\begin{figure}[ht]
\plotfiddle{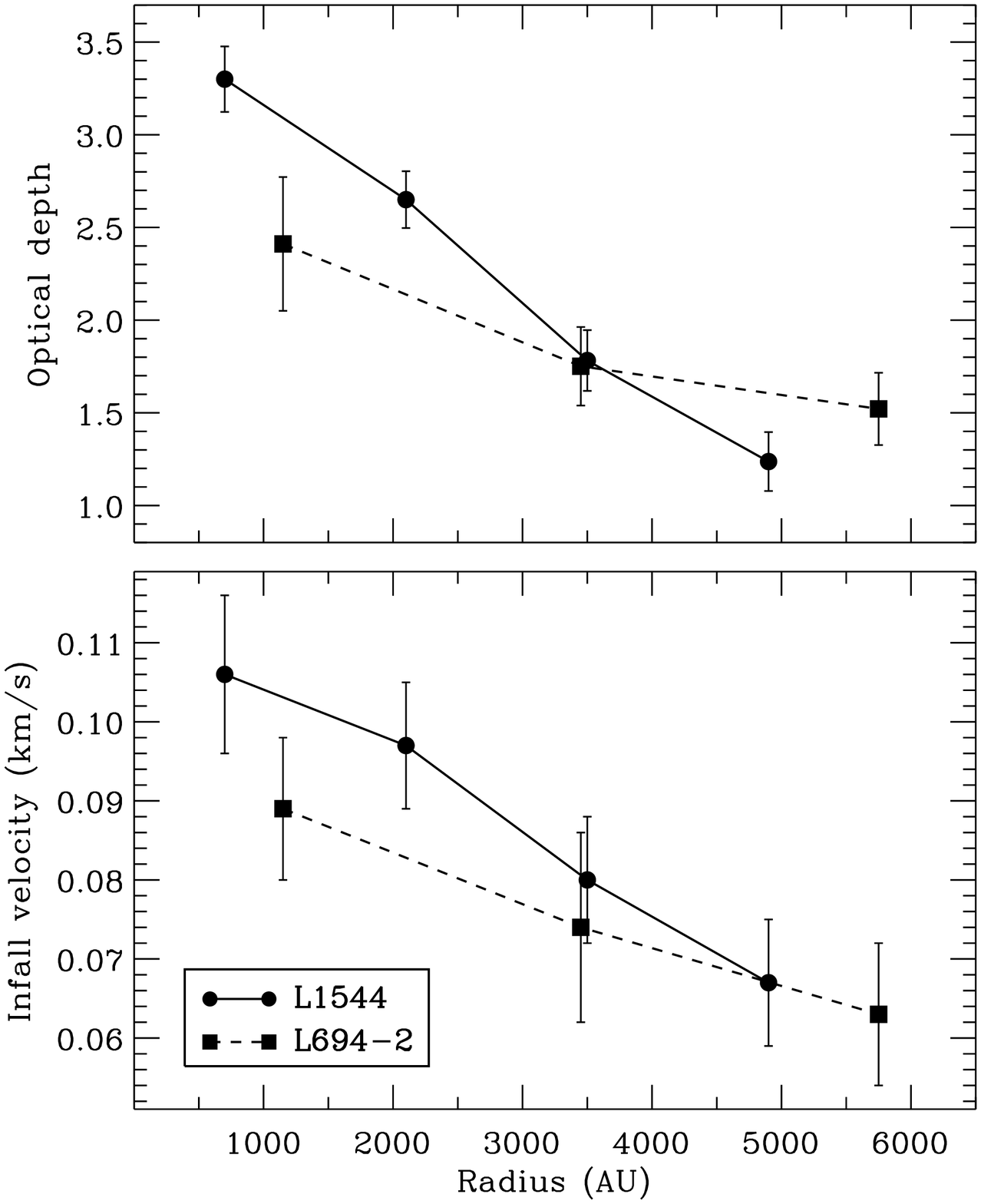}{50pt}{0}{50}{50}{-160}{-300}
\end{figure}
\vskip 3.7in
\noindent{\bf Figure 6:}
Optical depth and infall speed variation with radius,
derived from the two-layer spectral fits.
The errors were determined from multiple fits to fake
spectra created by adding noise to the best fit models.
L694-2 (squares connected by a dashed line) has a shallower
opacity gradient and lower central values of opacity
and infall speed than L1544 (circles connected by a solid line).
This may reflect an earlier evolutionary state or a slightly
lower mass.

\clearpage
\begin{table}
\begin{center}
TABLE 1\\
Two layer infall fit parameters\\
\vskip 2mm
\begin{tabular}{cccccc}
\hline\\[-2mm]
Annulus & $v_{LSR}$ & $\sigma$ & $T_{\rm peak}$ & $\tau$ & $v_{in}$ \\
 ($''$) &  (\kms)   &  (\kms)  &     (K)        &        &  (\kms)  \\[2mm]
\hline\hline\\[-3mm]
\multicolumn{6}{c}{\bf L694-2}\\[2mm]
$0-10$  &  9.61  &  0.078  &  9.3  & $2.4\pm 0.36$ & $0.089\pm 0.009$ \\
$10-20$ &  9.61  &  0.081  &  8.9  & $1.8\pm 0.21$ & $0.074\pm 0.012$ \\
$20-30$ &  9.61  &  0.081  &  8.0  & $1.5\pm 0.20$ & $0.063\pm 0.009$ \\[2mm]
\hline\\[-2mm]
\multicolumn{6}{c}{\bf L1544}\\[2mm]
$0-10$  &  7.20  &  0.084  &  7.1  & $3.3\pm 0.18$ & $0.106\pm 0.010$ \\
$10-20$ &  7.21  &  0.083  &  6.1  & $2.7\pm 0.15$ & $0.097\pm 0.008$ \\
$20-30$ &  7.22  &  0.086  &  5.8  & $1.8\pm 0.16$ & $0.080\pm 0.008$ \\
$30-40$ &  7.22  &  0.086  &  5.4  & $1.2\pm 0.16$ & $0.067\pm 0.008$ \\[2mm]
\hline
\end{tabular}
\end{center}
\end{table}

\end{document}